\documentclass[12pt]{iopart}
\usepackage{graphicx}
\begin{document}

\title[ ]{Relativistic Nucleus-Nucleus Collisions:
from the BEVALAC to RHIC}

\author{Reinhard Stock}

\address{University of Frankfurt/Main, Germany}

\begin{abstract}
I briefly describe the initial goals of relativistic nuclear
collisions research, focusing on the LBL Bevatron/Bevalac facility
in the 1970's. An early concept of high hadronic density fireball
formation, and subsequent isentropic decay (preserving information
as to the high density stage) led to an outline of physics
observables that could determine the nuclear matter equation of
state at several times nuclear ground state matter density. With
the advent of QCD the goal of locating, and characterizing the
hadron-parton deconfinement phase transformation suggested the
need for higher $\sqrt{s}$, the research thus moving to the BNL
AGS and CERN SPS, finally to RHIC at BNL. A set of physics
observables is discussed where present data span the entire
$\sqrt{s}$ domain, from Bevalac and SIS at GSI, to top RHIC
energy. Referring, selectively, to data concerning bulk hadron
production, the overall $\sqrt{s}$ evolution of directed and
radial flow observables, and of pion pair Bose-Einstein
correlation are discussed. The hadronization process is studied in
the grand canonical statistical model. The resulting hadronization
points in the plane T vs.~$\mu_B$ converge onto the parton-hadron
phase boundary predicted by finite $\mu_B$ lattice QCD, from top
SPS to RHIC energy. At lower SPS and top AGS energy a steep
strangeness maximum occurs at which the Wroblewski parameter
$\lambda_s \approx $ 0.6; a possible connection to the QCD
critical point is discussed. Finally the unique new RHIC physics
is addressed: high $p_T$ hadron suppression and jet "tomography".
\end{abstract}



\maketitle

\section{Introduction: Bevalac Physics}

In the early 70's a group of about 30 physicists settled at the
LBL Bevatron-Bevalac facility to start exploitation of
relativistic nucleus-nucleus collisions at fixed target energies
ranging up to 2 GeV per projectile nucleon. This group consisted
of scientists from the LBL Nuclear Science Division (both
experimental and theoretical) as well as from the Physics and
Accelerator Divisions, with significant migration from Germany and
Japan. With pioneering theoretical work by the Frankfurt,
Livermore and Los Alamos groups~\cite{1,2,3} the central goal was
to create "shock compression" in extended volumes of nuclear
matter, promising an avenue to investigate the properties of
compressed baryonic or, more generally, hadronic matter - at
densities several times the nuclear matter ground state density
$\varrho_0$ of 0.15 baryons per fm$^3$ or, equivalently, at energy
density $\epsilon_0 \approx$ 0.14 GeV per fm$^3$. Concurrent
astrophysics theory~\cite{4,5} had indicated that collective
hadronic matter properties (such as compressibility, temperature
and entropy density) at densities reaching up to about $4 \:
\varrho_0$ were required to understand type 2 supernova dynamics
as well as its remnants: neutron stars. All this information was
perceived to be contained in the equation of state (EOS) of high
density hadronic matter which relates pressure to hadronic number
density and temperature (in short: to the hadronic energy
density).

\begin{figure}
\begin{center}
\includegraphics[width=9cm]{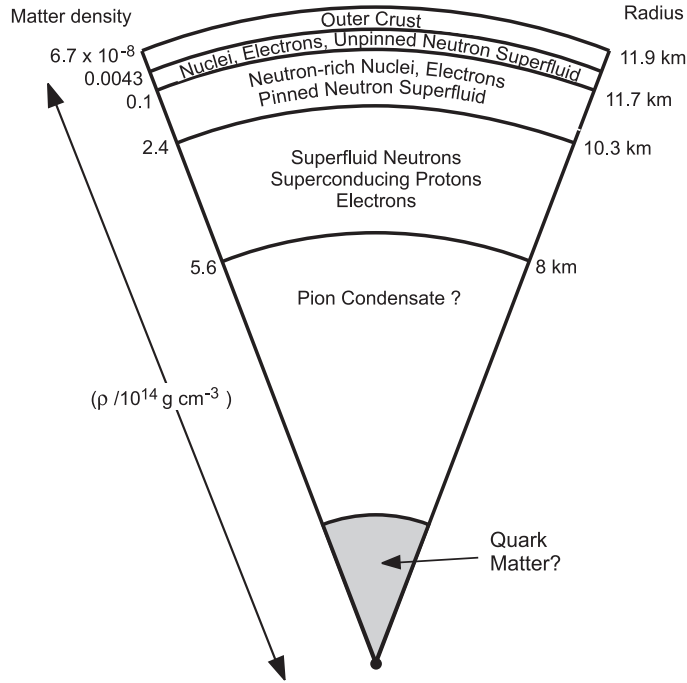}
\end{center}
\caption{The radial density profile of neutron stars reflecting
the hadronic EOS and the general QCD phase diagram.} \label{fig1}
\end{figure}

The central role of the hadronic EOS is illustrated by
Fig.~\ref{fig1} which shows the radial density profile of a
neutron star (of about 1.4 solar mass) from about 1975
astrophysics theory vintage. Under gravitational compression the
hadronic density increases toward the interior sections.
Hydrostatic equilibrium demands that, at any radial shell, the
gravitational inward pressure of the matter above is balanced by
the incompressibility outward pressure exerted by the matter
below. As the radial scale parameter is the density in this
diagram one needs the nuclear matter pressure vs.~density
relationship, i.e. the EOS, to solve for stability. The
hypothetical density diagram of Fig.~\ref{fig1} suggests,
furthermore, that the effective degrees of freedom of hadronic
matter (at T $\approx$ 0 characteristic for the neutron star)
should change with increasing density, thus affecting the overall
pressure to density relationship once various conceivable
collective effects (superfluidity, superconductivity, meson
condensates) set in. At the limit of cold hadronic matter
compression in the far interior, Fig.~\ref{fig1} includes a
further potential phase that stemmed from early QCD
studies~\cite{6}: a deconfined state of quarks, the "quark-gluon"
state of strongly interacting matter, obviously featuring a
radical change concerning the active  degrees of freedom that
dominate the overall EOS. In modern terminology it is, thus, the
general QCD phase diagram that underlies, as an example, neutron
star structure. The former is the subject of relativistic nuclear
collision research as we see it today.

\subsection{Bevalac observables}

Backward from the present formulation of research goals we see the
Bevalac research period devoted to an initial clarification of
physics observables that could elucidate  the high density
hadronic matter EOS. At the cost of greatly over-simplifying the
experimental and intellectual lines of development several crucial
steps can be distinguished

\noindent The first data on proton, neutron, pion and deuteron
production in central Ne + heavy target collisions were understood
in the "Fireball-model"~\cite{7}. As an adaption of the Hagedorn
statistical hadron production model~\cite{8} to central nuclear
collisions, it stated that the initial longitudinal beam energy
gets trapped in the target-projectile geometrical overlap volume
which is highly excited and compressed: the participant-spectator
model (spectators are the heavy target nucleons not directly hit
by the incident, relatively light projectile nucleus, initially
$^{12}$C, $^{20}$Ne, $^{36}$Ar) as combined with Hagedorn
statistics concerning the fireball decay to asymptotically free
hadrons.

\noindent The critical question: Does this finally observed hadron
gas contain any information relevant to the primordial high energy
density fireball; or is any such information lost due to hadronic
rescattering plus entropy increase, supposed to occur during the
fireball expansion period? Two crucial ideas emerged:

\noindent Firstly, Bertsch and Cugnon~\cite{9} showed that the
hadronic expansion stage is essentially {\it isentropic} and,
thus, in principle information preserving. In particular, an
isentropic expansion mode lends itself to a hydrodynamical
description~\cite{3,10}, which was widely employed later on. And
it generates the directly observable radial flow signal (well
studied from Bevalac to RHIC) that was proposed by Rasmussen
et~al.~\cite{11}. In isentropic expansion the configuration space
volume increases but the momentum space volume has to shrink
commensurably, giving rise to dimensional reduction by developing
a radially ordered "blast wave" momentum pattern that reflects the
entire expansion history, onward from its initial phase.

\noindent Secondly, Mekjian~\cite{12} recalled the phenomenology
of explosive nucleosynthesis occuring in the first minute of the
cosmological "big bang" fireball expansion. The light nuclear
species formed in a high temperature environment  of initial
protons, neutrons and photons stabilizes in a rate equilibrium
between strong interaction binding to deuterium, helium etc., and
dissociation by high energy photons. Upon expansion and cooling
this light cluster population {\it freezes out} at a critical
temperature, travelling onward unaltered by softer collisions as
the universe expanded. Analogously, Mekjian predicted that the
final population of hadrons becomes stationary {\it early} in
nuclear fireball decay, thus capturing a high energy density stage
that freezes-out initially, surviving isentropic rescattering.
This early work led to the present method of grand canonical
statistical analysis of hadron compositions, revealing the energy
density "at birth" of the hadronic phase~\cite{13,14}.

\noindent Finally, the above indications of isentropic decay were
systematically employed in the hydrodynamical model~\cite{3,10}
leading to the prediction of collective "bounce-off" flow of
nuclear matter: the primordial geometrical distribution of high
density matter and, thus, the pressure distribution gives rise to
a {\it directed} expansion in semi-peripheral nucleus-nucleus
collisions which are asymmetric in azimuth. This effect was
observed with striking clarity by the GSI-LBL Plastic Ball
detector~\cite{15} which made it possible --- by its completely
exclusive, 4$\pi$ observation of all emitted hadrons --- to pin
down the event-by-event direction of the impact vector, then to
re-order the emitted hadron transverse momenta with respect to the
impact plane. Whereupon the direction of preferred emission (flow)
becomes visible.

\subsection{The Bevalac legacy}

What emerged was an overall scheme of large acceptance study that
captures a significant fraction of the total hadron production
output thus allowing for an event-by-event analysis, coupled to a
fixation of the eventwise orientation of the impact vector thus
leading to analysis relative to the reaction plane. TPC tracking
in large acceptance was improved stepwise, from the LBL EOS TPC in
the HISS spectrometer (that later went on to the AGS experiment
E895), to the CERN SPS experiments NA49 and NA45, onward to STAR
at RHIC and to ALICE at the CERN LHC which is under construction.
>From Bevalac to LHC the TPC granularity has increased by a factor
of about 60, and so has the track multiplicity per central
collision event. The size of our experimental collaborations has
grown by roughly the same factor, such that a general scaling law
arises, not of produced hadrons per participant baryon (which
increases, from Bevalac to LHC, by about 60) but of participating
scientists per produced hadron (which stays about constant).
Pleading for forgivingness in view of such loose observations, I
assume that the LHC ALICE Experiment, with about 800 participating
scientists, will be well-prepared to cope with a midrapidity
charged particle density of dN/dy $\leq$ 4000.

I conclude that the Bevalac physics era (1972 - 1984) has endowed
our expanding research field with a wealth of physics observables,
initially designed for pinning down the equation of state of dense
hadronic matter. As research expanded toward illucidating the
general QCD phase diagram, and moved onward to higher $\sqrt{s}$
at the AGS, SPS, and RHIC facilities, two principal lines of
investigation, as inherited from early Bevalac research , have
remained until today: one tries, firstly, to outline observables
that freeze-out at early times thus capturing various relevant
high density stages of the dynamical evolution. Secondly, one
investigates signals that build up over extended periods of the
dynamical evolution. At Bevalac times observables of the first
kind were seen in electron-positron pair spectroscopy (by the DLS
spectrometer collaboration~\cite{16}) which refers to the
interpenetration phase, in hadron production ratios that stabilize
at the end of the high density phase~\cite{12,13,14}, and by means
of two pion HBT interferomery which captures the system at the
late times when pions emerge from their last collision. From among
the integral signals one studied radial and directed flow plus a
first view of what became known lateron as elliptical
flow~\cite{17}. Theory confronted all these observables within the
hydrodynamical model~\cite{3,10,18}, and by means of first
microscopic transport models such as VUU~\cite{19}, which included
a time dependent mean field. The effort to pin down the nuclear
EOS which had, at first, indicated a rather stiff
version~\cite{20} turned into a certain crisis when it was
realized~\cite{21} that most of the apparent stiffness stems from
high momentum hardening of the effective nucleon-nucleon forces
which predominates in high T fireball dymanics but is absent in
neutron stars and supernovae. This topic found its conclusion in
subsequent studies conducted at the GSI SIS facility: a semi-soft
EOS emerged~\cite{22}.

\section{Hadronic observables vs.~$\sqrt{s}$: from Bevalac to RHIC}

I turn to a brief discussion of a few selected physics observables
for which comprehensive data are now available, spanning the
entire center of mass energy range from the Bevalac to RHIC, i.e.
2..6 $\leq \: \sqrt{s} \: \leq$ 200 GeV. By implication these are
mostly observables referring to bulk hadron production, and I thus
refrain from discussion of the equally important physics offered
by "penetrating probes", i.e.~dilepton and direct photon
spectroscopy.

\subsection{Elliptic Flow}

\begin{figure}
\begin{center}
\includegraphics[width=10cm]{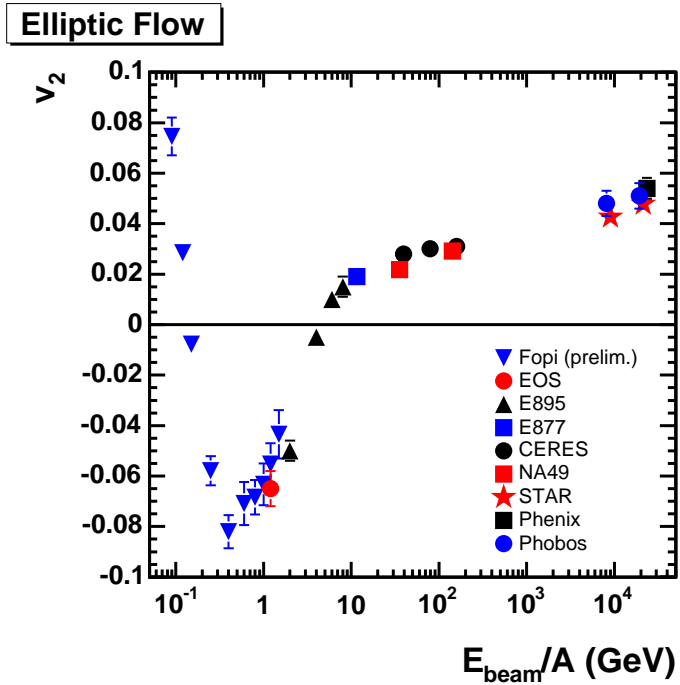}
\end{center}
\caption{Presently available data for the elliptic flow
coefficient $v_2$ (near midrapidity, integrated over $p_T$), for
$\sqrt{s}$ from SIS/Bevalac via AGS and SPS to RHIC.} \label{fig2}
\end{figure}

Heavy projectile collisions offer a high multiplicity of hadrons,
produced per event, that enables an event-wise study of the
azimuthal hadron emission pattern. By definition any azimuthal
emission anisotropy vanishes when the impact parameter $b$
approaches zero in central collisions, giving rise to approximate
cylindrical symmetry and maximal radially symmetric flow. However,
at finite $b$ we encounter the vector orientation of $b$ that
breaks cylinder symmetry of the hadron emission pattern. After
fixing this orientation (the event plane) event-by-event, the
first harmonic in the emission angle $\Theta$ with respect to this
plane essentially gives the "bounce-off" flow invented at the
Bevalac whereas the weight coefficient $v_2$ of the second
harmonic quantifies the elliptic flow. At high $\sqrt{s}$ it peaks
at $\Theta$=0 and 180$^0$, and it disappears toward
target/projectile rapidities as PHOBOS has shown~\cite{23}. It
arises from the gradient of the asymmetric pressure distribution
prevailing at the instant of maximal geometrical overlap of the
colliding nuclei and is, therefore, sensitive to the EOS of the
primordial collision volume~\cite{24}. Fig.~\ref{fig2} shows a
synopsis of all presently available data for $v_2$ (near
midrapidity, integrated over $p_T$), provided by
A.~Wetzler~\cite{25} for $\sqrt{s}$ from SIS/Bevalac via AGS and
SPS to RHIC. The initial, negative signal (from SIS/Bevalac to
lower AGS) reflects the fact that "cold" target-projectile
spectator matter is geometrically shadowing emission into the
direction of $\overline{b}$ so that only "side
splash"~\cite{17,18} is available to emission from the primordial
interaction volume (implicitly demonstrating that elliptic flow is
an early time emission process). This shadowing disappears at
Lozenz-$\gamma \ge 3$; where the v$_2$ signal turns positive. A
tantalizing hint at a saturation occurs toward top SPS, the rise
resuming at the two RHIC energies (one of the key arguments for
RHIC running at lower $\sqrt{s}$). This signal increases with
transverse momentum~\cite{26}, plausibly so as the collective,
anisotropic pressure field accelerates particles differently
in-plane and out-of plane --- an effect that is well reproduced by
hydrodynamics with partonic EOS~\cite{27} at RHIC energy.

\begin{figure}
\begin{center}
\includegraphics[width=8cm]{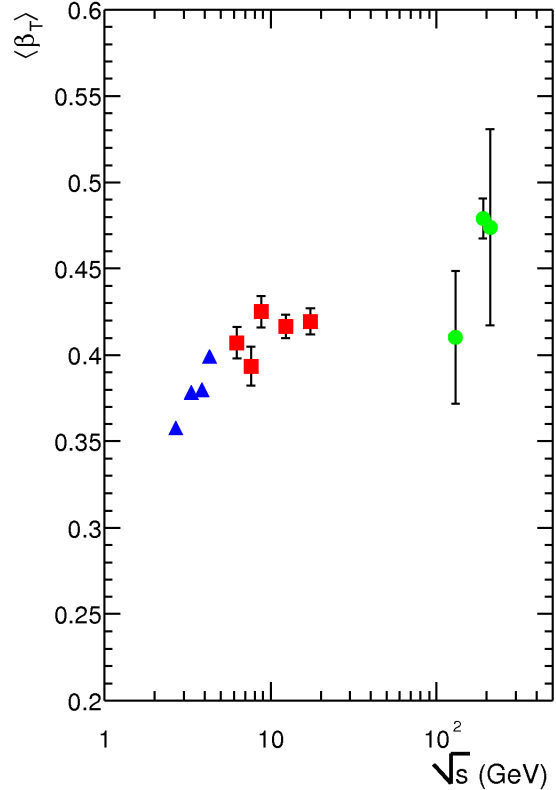}
\caption{The mean radial velocity $<\beta_T>$ as a function of
$\sqrt{s}$ from Bevalac to RHIC.} \label{fig3l}
\end{center}
\end{figure}

\subsection{Radial Flow and spectral "Temperature"}

\begin{figure}
\begin{center}
\includegraphics[width=14cm]{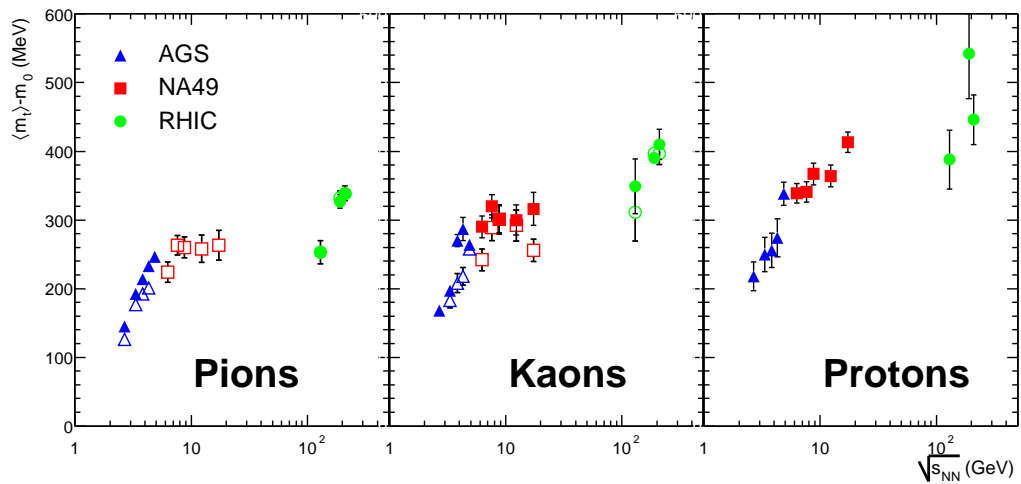}
\end{center}
\caption{Mean transverse momenta for pions, kaons and protons at
midrapitidy as a function of $\sqrt{s}$ indicating an intermediate
plateau in the SPS region.} \label{fig3m}
\end{figure}

\begin{figure}
\begin{minipage}[b]{0.5\linewidth}
\centering
\includegraphics[width=8cm]{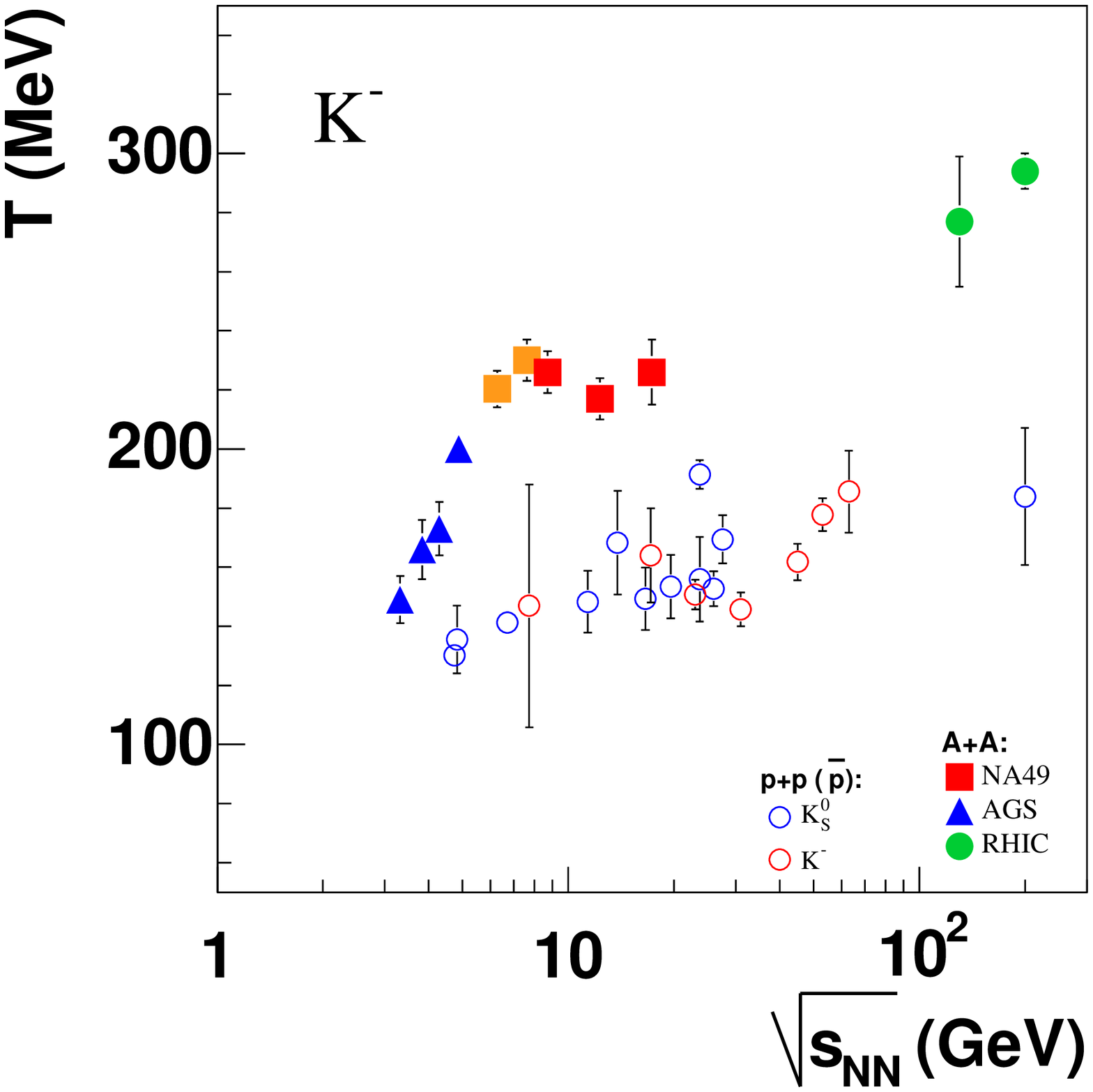}
\end{minipage}
\hspace{0.0cm} % To get a little bit of space between the figures
\begin{minipage}[b]{0.5\linewidth}
\centering
\includegraphics[width=8cm]{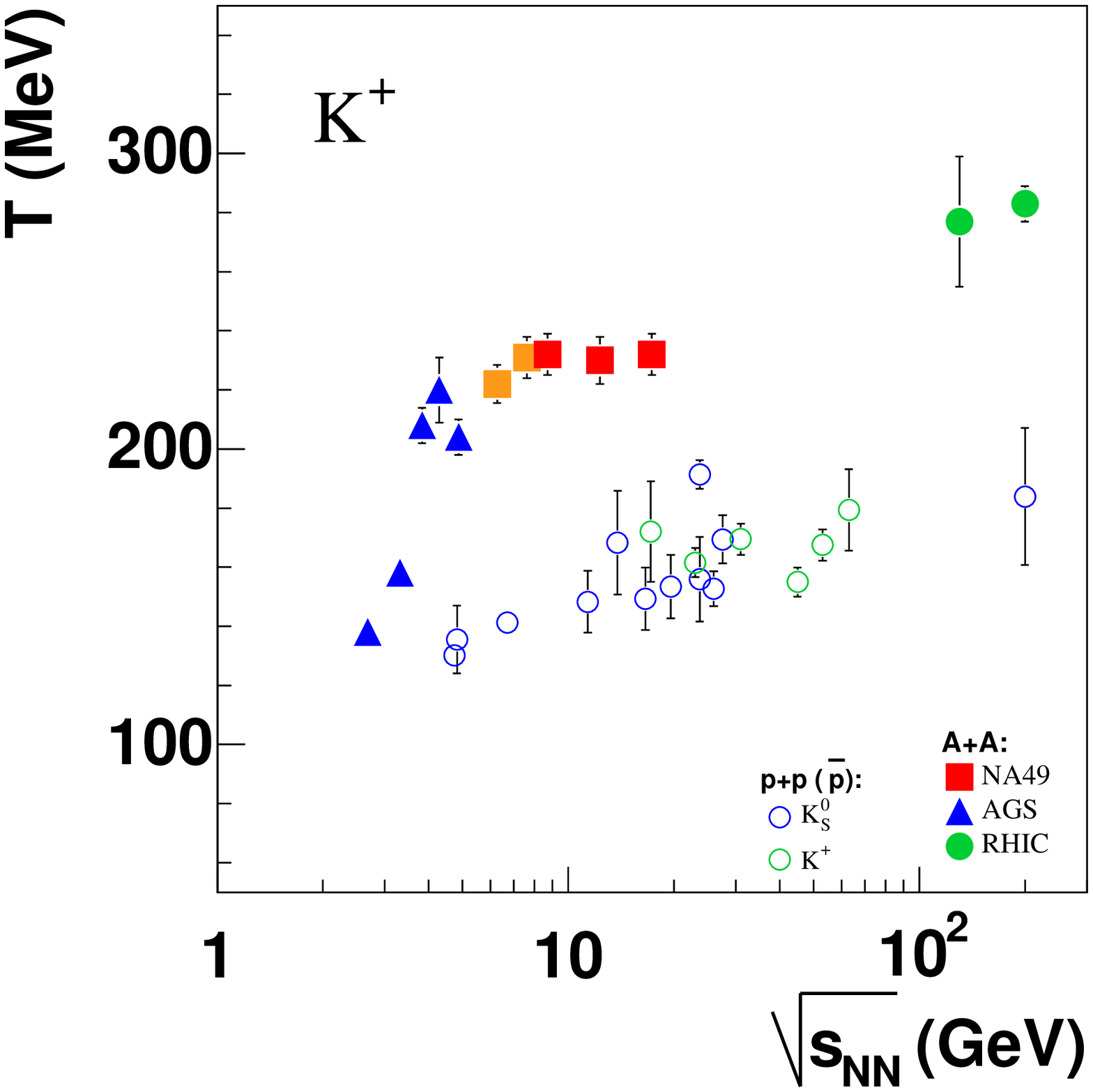}
\end{minipage}
\caption{The inverse slope parameter (the "spectral temperature")
for charged kaons as a function of $\sqrt{s}$ .} \label{fig3r}
\end{figure}

A radially symmetric momentum orientation arises upon isentropic
expansion of an initially dense system, be it partonic or
hadronic, or both ways in turn. Hydrodynamical or hydrodynamically
"inspired" models (which assume an isentropic expansion by
definition) describe this effect by a radial collective velocity
field that increases toward the surface, at each instant of time,
the surface (and average) velocity increasing over the entire
course of the expansion. Thus the flow fraction of the average
kinetic energy increases while the temperature falls steeply, both
observables reaching a certain characteristic value that
characterizes the stage where emission products decouple from
rescattering. From Bevalac to RHIC the mean value $<\beta_T>$
increases from about 0.35 to about 0.6~\cite{28} as is shown in
Fig.~\ref{fig3l}. After an initial steep rise a hint at saturation
is again observed within the SPS energy range, which is overcome
by the $\sqrt{s} = 130$ and $200$~GeV RHIC data points. Such an
intermediate plateau is also indicated by the $\sqrt{s}$
dependence of the midrapidity mean transverse momenta of various
hadronic species, a completely model-independent
observation~\cite{28,29} shown in Fig.~\ref{fig3m}. Finally I wish
refer to the well known "Nu-Xu" plots~\cite{30,31} which
illustrate the effect of the radial flow velocity field on the
inverse slope parameter of transverse mass spectra of various
hadronic species. This "spectral temperature" turns out to be more
influenced by the ordered flow velocity field than by the
remaining random thermal velocity. It thus exhibits an overall
increase with $\sqrt{s}$~\cite{32}. The temperature plot is shown
in Fig~\ref{fig3r} specifically for the $\sqrt{s}$ systematics of
charged kaons. The reason for this particular choice is that kaon
transverse mass spectra are known from Bevalac/SIS to RHIC to have
a nearly exponential shape, thus rendering themselves to a simple
inverse slope analysis --- unlike pions which feature concave
$p_T$ or $m_T$ spectra due to pronounced resonance decay
contributions, and also unlike baryons with their pronounced
"shoulder arm structure" at low and intermediate $p_T$ or $m_T$
(which requires two parameter fits by, e.g.~the modern blast wave
model~\cite{33}). Obviously, both charged kaon species exhibit a
pronounced structure of turning from initial rise into a plateau,
then into rising again at the two RHIC energies presently
available.

\noindent All these patterns (Figs.~2-5)remind one of a phase
diagram featuring a parton-hadron coexistence phase of QCD matter.
But there may be alternative explanations for such a $\sqrt{s}$
dependence and, in any case, one concludes that RHIC runs at
intermediate $\sqrt{s}$ are highly desireable to confirm such
indications of a plateau structure.

\begin{figure}
\begin{center}
\includegraphics[width=10cm]{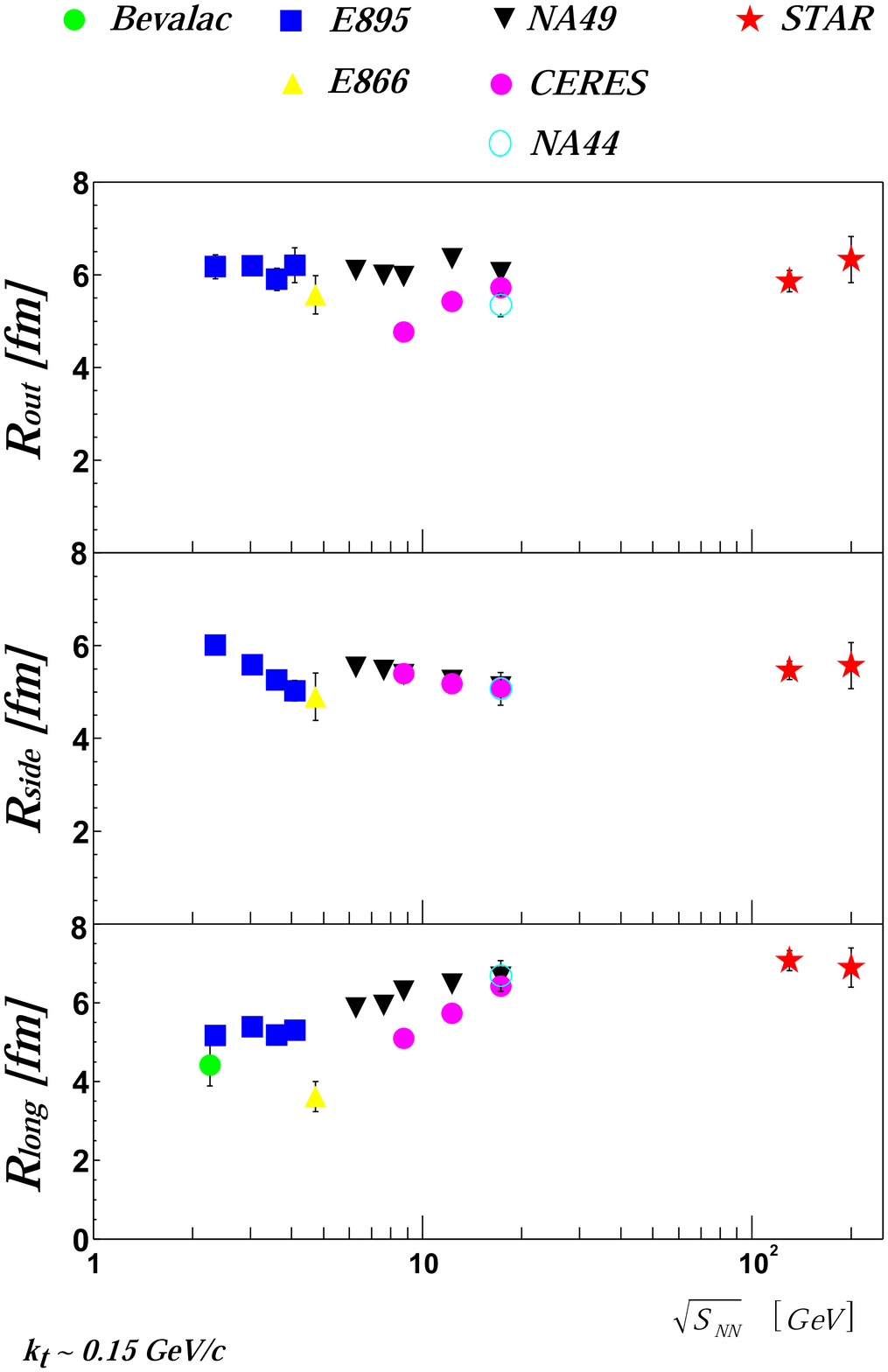}
\end{center}
\caption{Compilation of HBT parameters~\cite{35} derived for
positive and negative pion pair production as a function of the
$\sqrt{s}$.} \label{fig6}
\end{figure}

\subsection{Identical Pion Bose-Einstein Correlation}

I turn to pion Bose-Einstein correlations in order to introduce a
certain sense of caution concerning the high expectations invited
by the above discussion of collective flow signals. From the
theoretical models~\cite{38} developed in the 1990's utilizing a
wide spectrum of hydrodynamical approximations concerning the
space-time-momentum evolution of the expanding system one expects
a multitude of characteristic features which are mostly absent in
the data. Fig.~\ref{fig6} shows a compilation~\cite{39} of the
$\sqrt{s}$ dependence of HBT parameters derived for positive and
negative pion pair production observed at low pair momenta in
central collisions of heavy nuclei.The 3 characteristic
"geometrical" parameters of the Pratt-Bertsch model~\cite{38},
$R_{out}$, $R_{side}$ and $R_{long}$ are, essentially, constant
and equal to each other, of order 6 fm with exception of the
Bevalac point which, however, refers to $^{38}$Ar as a projectile.
This deviation from intuitively expected features (such as radii
growing with the square or cube roots of d$N$/d$y$, $R_{out}$
significantly exceeding $R_{side}$ etc.) may stem from a radically
wrong picture, employed in all hydro-typ models, concerning the
pionic decoupling stage or, more precisely, the hypersphere in
space-time for pionic freeze-out. This stage may differ from naive
pictures concerning pion freeze-out (see H.~Appelsh\"auser talk at
this conference~\cite{40}), or stem more radically, from an onset
of "instantaneous" hadronic freeze-out toward higher
$\sqrt{s}$~\cite{41} that is not yet well captured by dynamical
models.

\section {Statistical Hadron Production}

Bulk hadron production systematics in central nucleus-nucleus
collisions at relativistic energy is, overall, well reproduced by
a statistical Hagedorn hadronic freeze-out model. A grand
canonical version of this model captures the various hadronic
species multiplicities, per collision event, from pions to
$\Omega$~hyperons, in terms of a few universal parameters that
describe the dynamical stage in which the emerging hadronic matter
decays to a quasi-classical gas of free resonances and
hadrons~\cite{13,14,38,39}. The grand canonical parameters are
temperature~$T$, volume~$V$ and chemical potential~$\mu$. They
capture a snapshot of the fireball expansion within the narrow
time interval surrounding hadronic chemical freeze-out, which thus
appears to populate the hadron/resonance mass and quantum number
spectrum, predominantly, by phase space weight~\cite{8,14,40} thus
creating an apparent thermal equilibrium state prevailing in the
produced hadron-resonance-population. This chemical equilibrium
instantaneously decouples from fireball expansion surviving
further (near isentropic) processes. It can thus be retrieved from
the finally observed hadronic multiplicities, by state of the art
grand canonical model analysis. This analysis succeeds from AGS,
via SPS, to RHIC energy.

Statistical model analysis is also applicable to elementary
collisions, $p+p$, $p+\overline{p}$, and $e^+e^-$ annihilation as
was shown by Hagedorn~\cite{8} and, more recently, by Becattini
and Collaborators~\cite{41,42}. The canonical version of ensemble
analysis is applicable here. Mutatis mutandis the same
hadrochemical equilibrium feature is being attested, emphasizing
the statement that the apparent equilibrium does not arise from an
inelastic rescattering cascade toward thermodynamical equilibrium
- there is essentially none in elementary processes - but should
stem directly from the QCD hadronization process occuring under
phase space dominance~\cite{14,40}.

The crucial difference between elementary and central
nucleus-nucleus collisions resides, in statistical model view, in
a transition from canonical to grand canonical order in the
ensuing decoupled hadronic state. This transition was studied by
Cleymans, Tounsi, Redlich et~al.~\cite{43}. Its main feature is
strangeness enhancement. Comparing the strange to non-strange
hadron multiplicities in elementary, and in central
nucleus-nucleus collisions at similar energy, one observes an
increase of the singly strange hyperons and mesons, relative to
pions, of about 2-4, and corresponding higher relative
enhancements of multiply strange hyperons~\cite{44,45,46}, ranging
up to order-of-magnitude enhancement. In the terminology of
Hagedorn statistical models, strangeness is suppressed in the
small system, canonical case, of elementary collisions (due to the
dictate of local strangeness conservation in a small "fireball"
volume), whereas it approaches flavour equipartition in large
fireballs due to the occurence of quantum number conservation, on
average only, over a large volume
--- as reflected by the {\bf global} chemical potential featured by
the grand canonical ensemble: "strangeness enhancement" occurs as
the fading-away of canonical constraints.

>From statistical model analysis we obtain a more general view of
strangeness relative to non-strangeness production than is
provided by considering individual strange to non-strange
production ratios, like $K/\pi, \:\Omega/\pi$ etc., from $p+p$ to
central A+A. The model quantifies strange to non-strange
hadron/resonance production by means of Wroblewski quark counting
at hadronic freeze-out~\cite{47}. It determines the so-called
Wroblewski-ratio,
\begin{equation}
\lambda_s=\frac{2(<s>+<\overline{s}>)}{<u>+<\overline{u}>+<d>+<\overline{d}>}
\end{equation}

\noindent
 which quantifies the overall strangeness to
non-strangeness ratio at hadronic freeze-out. Strangeness
enhancement (i.e.~removal of strangeness suppression in elementary
collisions) is quantified, by such an analysis, to proceed from
$\lambda_s \approx 0.25$ in elementary collisions, to $\lambda_s
\approx 0.45$ in central nucleus-nucleus
collisions~\cite{13,38,39}.

\begin{figure}
\begin{center}
\includegraphics[width=9cm]{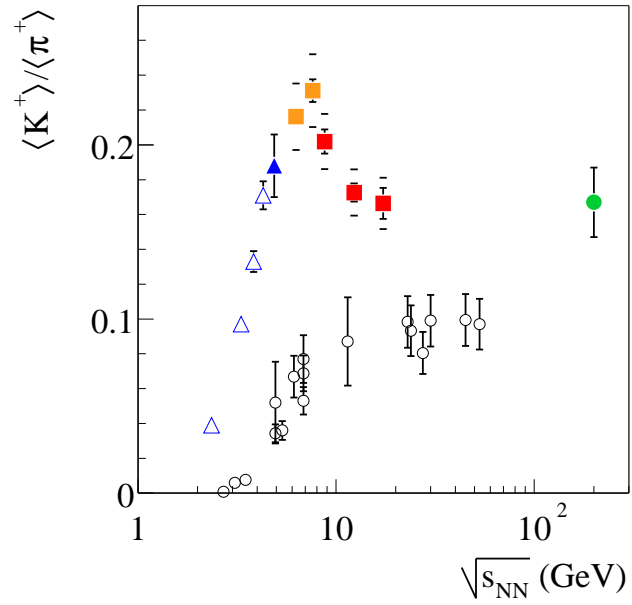}
\end{center}
\caption{Energy dependence of the $<K^+>/<\pi^+>$ ratio for
central Pb+Pb (Au+Au) collisions (upper points) and p+p
interactions (lower points).} \label{fig5a}
\end{figure}

\begin{figure}
\begin{center}
\includegraphics[angle=270,width=80mm]{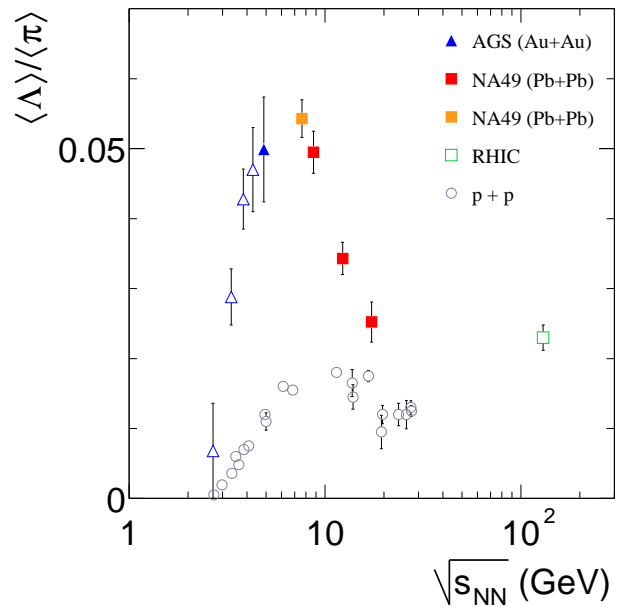}
\end{center}
\caption{ Energy dependence of the $<\Lambda>/<\pi>$
\mbox{($<\pi>=1.5\:(<\pi^->+<\pi^+>)$)} ratio for central Pb+Pb (Au+Au)
collisions (upper points) and p+p interactions (lower points).}
\label{fig5b}
\end{figure}

>From a recent energy scan conducted at the SPS by NA49, studying
hadron multiplicities from $\sqrt{s}=7$ to $17$~GeV, a steep
maximum was observed~\cite{48} in the $K^+/\pi$ and $\Lambda/\pi$
ratios in central Pb+Pb collisions, as shown in Figs.~\ref{fig5a}
and~\ref{fig5b}. As the $K^+$ and $\Lambda$ channels carry most of
the total $<s>+<\overline{s}>$ content, this experimental result
indicates a kind of "singularity" in the strange to non-strange
production ratio, from AGS to RHIC energy. This effect is absent
in $p+p$ collisions. Becattini et al.~\cite{39} analyzed the
$\sqrt{s}$ dependence of the Wroblewski-parameter $\lambda_s$ in
the grand canonical statistical hadronization model. Their result
is shown in Fig.~\ref{fig6} which gives $\lambda_s$ as a function
of the chemical potential~$\mu_B$. From top AGS energy (at $\mu_B
\approx 550$~MeV) to RHIC energy ($\mu_B \leq 50$~MeV) one
perceives an average $\lambda_s$ of about $0.45 \pm 0.08$ whereas
a steep excursion is seen, to $\lambda_s=0.6 \pm 0.1$, at
$\mu_B=440$~MeV. This point corresponds to the steep maxima
observed in Figs.\ref{fig5a} and \ref{fig5b}, to occur at SPS
fixed target energy of 30 GeV/A in central Pb+Pb collisions,
corresponding to $\sqrt{s}=7.3$~GeV.  The observed $\lambda_s$
maximum should, therefore, present a hint that hadronization at
$\sqrt{s} \approx 7$~GeV should occur under influences, absent at
energies above and below. Moreover, NA49 has shown
recently~\cite{49} that the event-by-event fluctuation of the
ratio ($K^+ + K^-)/(\pi^+ + \pi^-$) measured in central Pb+Pb
collisions increases steeply toward $\sqrt{s} = 7$~GeV whereas it
was formerly found~\cite{50} to amount to be below 4\%, at top SPS
energy, $\sqrt{s} = 17.3$~GeV.

\begin{figure}
\begin{center}
\includegraphics[width=9cm]{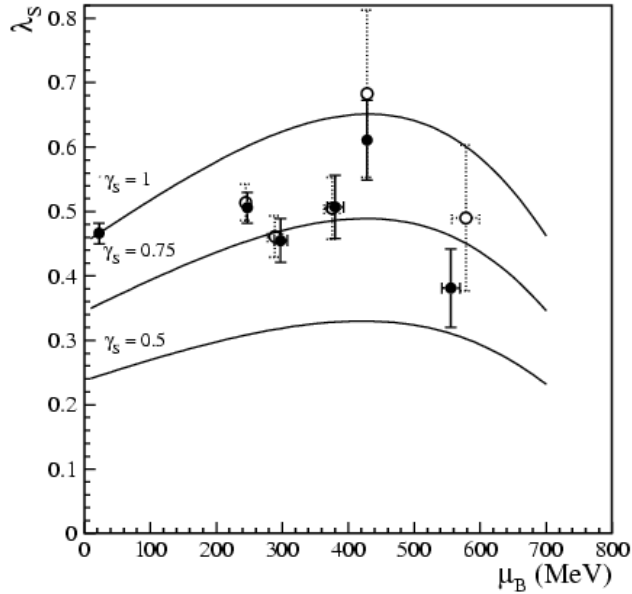}
\end{center}
\caption{Dependence of the $\lambda_s$ parameter on baryochemical
potential extracted from the fits to hadron multiplicities in
central Pb+Pb (Au+Au) collisions at AGS, SPS and RHIC energies.
The lines show the dependence expected for different values of the
$\gamma_s$ parameter [ref.~39]} \label{fig9}
\end{figure}

We thus propose that the dynamical trajectory of central Pb+Pb
collisions comes close to the critical point of QCD, at or near
$\sqrt{s} = 7$~GeV. This point has been expected to occur on the
line in the $T$, $\mu_B$ plane which describes the boundary
between the hadronic and partonic QCD phases~\cite{51}. Along that
line the phase transition is expected to be a crossover at
$\mu_B<\mu_B^c$, become second order at $\mu_B = \mu_B^c$, and
first order for $\mu_B > \mu_B^c$. At $\mu_B = \mu_B^c$ and $T T_c$ we thus expect phenomena analogous to critical opalescence.
Recent QCD lattice calculations succeeded in an extrapolation to
finite chemical potential~\cite{52,53}, thus making a first
prediction for the phase boundary line and, in particular, the
critical point - albeit with considerable uncertainty as was
discussed by Redlich at this conference~\cite{54}. This
uncertainty stems, firstly, from the uncertainty in the
extrapolation to finite $\mu_B$ but, secondly, from the unphysical
(high) strange quark mass employed in these lattice calculations
which, at present, place the critical point somewhere in the
interval 500~MeV $ < \mu_B^c \le 700$~MeV. Redlich argued that it
should move to considerably lower $\mu_b$ once the s-mass can be
chosen closer to the physical quark mass. This expectation was
substantiated by recent lattice calculations which show that the
critical point might move downward in $\mu_B$ once more realistic
quark masses are employed~\cite{55}. From Fig.~\ref{fig9} we see
that the strangeness maximum at $\sqrt{s} = 7$~GeV corresponds to
$\mu_B \approx 440$~MeV and thus quite close to the expected
$\mu_B^c$ position.  The energy density at the phase boundary is
estimated by lattice QCD to be rather low~\cite{56} ($\epsilon
\le$ 1 GeV/fm$^3$).

Central collisions of heavy nuclei at moderately relativistic
energy exhibit a general cycle of initial compression and heating
which is followed by a maximum energy density stage which then
turns into expansion and cooling~\cite{20}. The quantities
characterizing the overall system dynamics, such as volume and
energy-entropy density etc.~change very rapidly except during the
high density stage which acts analogous to a classical turning
point. If it coincides closely with the QCD critical endpoint one
could expect to observe substantial critical phenomena. Now it is
well known that the maximum energy density in central collisions
of mass 200 nuclei amounts (Bjorken estimate) to above 2
GeV/fm$^3$ at top SPS \cite{57}, and to about 5 GeV/fm$^3$ at RHIC
energies~\cite{58}, thus overshooting, by far, the critical QCD
energy density. The system thus crosses the phase coexistence
line, upon re-expansion, whilst already undergoing rapid
expansion. Furthermore, the chemical potential is certainly well
below 300 MeV at the time of hadronization. The evolution will
thus miss the critical point at top SPS, and RHIC energies; and at
much lower AGS energies the dynamics falls into the $\mu_B \ge
500$~MeV domain but the energy might not suffice to reach the
phase boundary. In summary we may indeed expect that the dynamical
evolution reaches its energy density plateau phase near the
expected critical point (i.e.~at energy density slightly below
1~GeV/fm$^3$, and at $\mu_B$ between 300 and 500~MeV) somewhere in
the domain of maximum AGS and minimum SPS energy.

\begin{figure}
\begin{center}
\includegraphics[width=9cm]{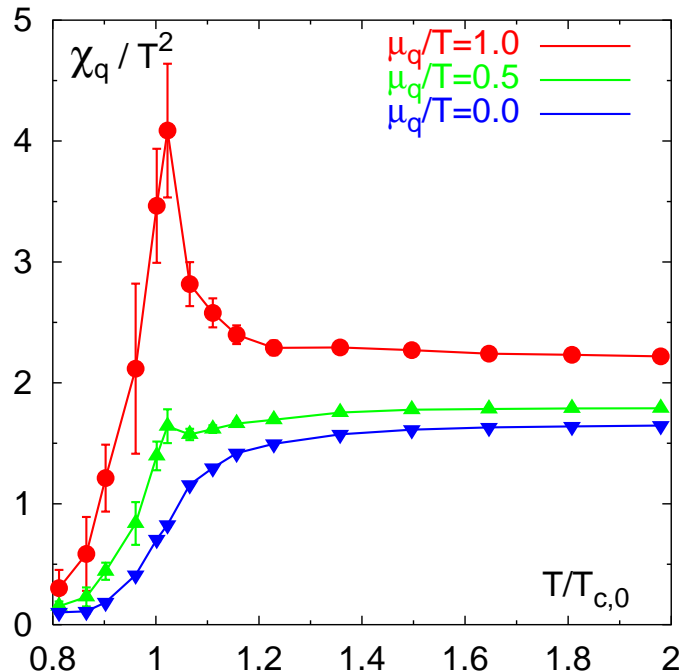}
\end{center}
\caption{The quark number susceptibility calculated within lattice
QCD as function of temperature (relative to transition
temperature) for different values of quark chemical potential.}
\label{fig7}
\end{figure}

Returning to the lattice results~\cite{53,54,55} at finite
$\mu_B$, we see a steep maximum of the quark number susceptibility
\begin{equation}
\chi_{u,d} \equiv T^2(\frac{d^2}{d(\mu/T)^2} \: \frac{p}{T^4})
\end{equation}

\noindent occuring at $T=T_c= 150$~MeV and $\mu_B = 3 \mu_Q = 3 T 450$~MeV. We reproduce the Bielefeld-Swansea results in
Fig.~\ref{fig7} which also shows the calculations
for $\mu_B = 225$~MeV and $\mu_B = 0$
(essentially corresponding to top SPS and RHIC energies,
respectively). The latter exhibit no susceptibility peak but a
smooth transition from $T < T_c$ to $T  >  T_c$. As
$\chi_{u,d}$ can also be written as
\begin{equation}
\chi_q = T^2(\frac{\delta}{\delta(\mu_{u/T})} \:+ \:
\frac{\delta}{\delta(\mu_{d/T})}) \: \frac{n_u + n_d}{T^3}
\end{equation}
we see that the peak in the susceptibility implies a maximum
fluctuation of the quark number densities $n_u$ and $n_d$. We
interpret this result as an indication of critical fluctuation
occuring in the vicinity of the critical endpoint implicitly
present in this calculation. Directly at $\mu_c$ the
susceptibility would diverge. The critical point in this
calculation must thus be near $\mu_B = 450$~MeV and $T = 150$~MeV.
This, in turn, is very close to the parameters of the grand
canonical model at the strangeness maximum, $\sqrt{s} = 7$~GeV
(Fig.~\ref{fig9}.

As to the relation between the susceptibility maximum of lattice
QCD and the strangeness maximum observed by NA49 (at which the
Wroblewski parameter $\lambda_s$ exhibits an anomaly), Gavai and
Gupta~\cite{59}  have suggested the relationship
\begin{equation}
\lambda_s = \frac{2 \chi_s}{\chi_u+\chi_d}
\end{equation}
which appears to offer a direct link. In fact they obtain
$\lambda_s = 0.48$ from a lattice calculation at zero $\mu_b$:
closely coinciding with the value observed at top SPS and RHIC
energy (Fig.~\ref{fig9}). Unfortunately, though, their result
refers to $\mu_b = 0$, and the Bielefeld-Swansea calculations at
finite $\mu_b$~\cite{53,55} are in two-flavour QCD only. A
prediction for $\chi_s$ at $\mu_B \approx 450$~MeV, or, more
generally, a full three-flavour lattice treatment of the vicinity
of the critical point is required to finally assess the above
argument that tries to establish a link between the Wroblewski
maximum and the parton density fluctuations at the critical point.

\section{New RHIC Physics: In-Medium Parton Attenuation and Jets}

It has been the point of the previous two chapters to present
observables for which we have comprehensive data from Bevalac/SIS
to RHIC. With the exception of HBT "radii" all these $\sqrt{s}$
dependences exhibit an interesting structure, of an initial steep
rise up to the lower SPS energies that is followed by indications
of a plateau extending over the SPS domain $7 \le \sqrt{s} \le
17$~GeV, then by a further steep rise occuring at RHIC energies
$\sqrt{s}=130$ and 200~GeV. This overall pattern has not yet been
theoretically understood. Qualitatively one might argue that the
plateau structure signals phase coexistence setting in over the
SPS energy range, whereas partonic, primordial dynamics becomes
dominant at top RHIC energies.

Before turning to this physics I wish to note, however, that the
two characteristic temperatures that describe the bulk hadronic
dynamical trajectory do indeed reach saturation from top SPS to
RHIC energy. The hadronization temperature (inferred from the
grand canonical statistical hadronization model) saturates at
about $165 \pm 10$~MeV, in close agreement with lattice QCD
estimates of the parton-hadron phase coexistence
domain~\cite{13,14,39}. Furthermore, the second characteristic
"freeze-out" temperature, that describes the final bulk hadronic
decoupling from strong interaction, appears to saturate at $100
\pm 10$~MeV. For lack of space I can not discuss here the possible
excursions from this universal picture of hadronic phase
expansion, as implied by hyperon data. Within such reservation the
{\it thermal} history of bulk hadron expansion may well turn out
to be universal from SPS to RHIC, while its hydrodynamical
parameters (radial and elliptic flow, as well as mean $p_T$ of
$m_T$) reflect the increasing influence of the pre-hadronic phase,
setting the stage for the ensuing hadronic expansion phase.

All of the above discussion has focused, implicitly, on low $p_T$
physics. The radically new physics, offered by RHIC, stems from
expanding our view to $p_T$ up to about 15~GeV. With RHIC we thus
turn from soft to hard QCD physics, approaching a situation in
which observed high transverse momentum hadrons stem from
primordial hard partonic rescattering as described by perturbative
QCD. The qualitatively new feature  (added to well known pQCD
hadron production as it gets imbedded into a large primordial
interaction volume) is the in-medium attenuation of the leading
partons that are initially emitted in a hard partonic scattering.
A colour charge propagating through a colour charged medium
suffers induced radiative energy loss thus modifying the well
known DGLAP evolution that describes leading parton hadronization
in vacuum (in elementary collisions). This energy loss is a QCD
analogy to the Landau-Pomeranchuk phenomena of QED that occur once
an electric charge traverses an extended electrically charged
medium. Due to quantum mechanical interference the net in-medium
radiative energy loss becomes proportional to the square of the
path-length $L$ over which the propagating charge interacts
in-medium. A complication arises as we are not dealing with a
homogeneous, infinite volume in central nucleus-nucleus
collisions: the simple L$^2$ law of radiative energy loss gets
modified by a transport coefficient that reflects the changing
local energy density as experienced by the leading parton
traversing the primordial fireball volume while it
expands~\cite{60}.

A vision emerges, for an experimental program of the RHIC
experiments (and for further studies at the CERN LHC) that could
verify the basic L$^2$ law of leading parton in-medium energy
loss. As a first step the RHIC experiments have demonstrated a
universal in-medium suppression of high $p_T$ bulk hadron
production yields, as compared to elementary collisions at similar
$\sqrt{s}$. The next step results from analysis of the
back-to-back production pattern of jets. This primordial hard
parton scattering signal offers a distinct geometrical pattern
which can be related to the overall geometry of an A+A collision
fireball, into which it is embedded. If a jet is created in the
periphery of the primordial fireball one of the emerging leading
partons may escape into free space essentially unattenuated
whereas its opposite side partner traverses the entire radial
extent of the reaction volume, thus being maximally attenuated.
This opposite side jet quenching phenomenon, as observed by the
RHIC experiments, can be quantified, both, versus reaction
centrality, and with respect to the location of the impact plane.
Such an analysis promises to unravel the two essential parameters
of jet attenuation study: the length L of the opposite side jet
parton traversal through the dense medium, and the integral of the
QCD transport coefficient over the entire trajectory of the
emerging opposite-side jet. The first quantity is merely
geometric, the second depends on a model of the radial energy
density distribution of the primordial fireball, and its evolution
during the time interval sampled by the opposite side leading
parton while fragmenting into an eventually observed jet.
Actually, a multitude of contributions to this Conference show
that at RHIC jet energies, in the domain of about 10 GeV studied
thus far, the opposite side jet is entirely quenched in central
Au+Au collisions, whereas it gradually appears toward smaller L as
encountered in semi-peripheral collisions. These, and expected
further RHIC data may thus result in verification of the QCD L$^2$
law, characteristic of a deconfined medium.

\end{document}